\documentclass[aps,pra,reprint,superscriptaddress]{revtex4-2}

\usepackage{graphicx}
\usepackage{dcolumn}
\usepackage{braket}
\usepackage{bm}
\usepackage{hyperref}
\usepackage{xcolor}
\usepackage{svg}
\usepackage{subfigure}
\usepackage{tabularx}
\usepackage{multirow}

\begin{document}

\preprint{APS/123-QED}

\title{Satellites promise global-scale quantum networks}

\author{Sumit Goswami}
\affiliation{Institute of Atomic and Molecular Sciences, Academia Sinica, Taipei City, Taiwan}
\affiliation{Institute for Quantum Science and Technology, and Department of Physics \& Astronomy, University of Calgary, 2500 University Drive NW, Calgary, Alberta T2N 1N4, Canada}

\author{Sayandip Dhara}
\affiliation{Virginia Tech, Blacksburg, VA 24061, United States}

\author{Neil Sinclair}
\affiliation{John A. Paulson School of Engineering and Applied Science, Harvard University, 29 Oxford St., Cambridge, MA 02138, USA}
\affiliation{Division of Physics, Mathematics and Astronomy, and Alliance for Quantum Technologies (AQT), California Institute of Technology, 1200 E. California Blvd., Pasadena, California 91125, USA}

\author{Makan Mohageg}
\affiliation{Boeing Research and Technology, 929 Long Bridge Drive, Arlington, VA 22202, USA}

\author{Jasminder S. Sidhu}
\affiliation{SUPA Department of Physics, University of Strathclyde, Glasgow G4 0NG, UK}

\author{Sabyasachi Mukhopadhyay}
\affiliation{Centre for Computational and Data Sciences, IIT Kharagpur, Kharagpur, West Bengal 721302, India}

\author{Markus Krutzik}
\affiliation{Institut für Physik and Center for the Science of Materials Berlin (CSMB), Humboldt-Universität zu Berlin, Newtonstr. 15, Berlin 12489, Germany}
\affiliation{Ferdinand-Braun-Institut (FBH), Gustav-Kirchoff-Str. 4, Berlin 12489, Germany}

\author{John R. Lowell}
\affiliation{Boeing Research and Technology, 929 Long Bridge Drive, Arlington, VA 22202, USA}

\author{Daniel K. L. Oi}
\affiliation{SUPA Department of Physics, University of Strathclyde, Glasgow G4 0NG, UK}

\author{Mustafa Gündoǧan}
\affiliation{Institut für Physik and Center for the Science of Materials Berlin (CSMB), Humboldt-Universität zu Berlin, Newtonstr. 15, Berlin 12489, Germany}

\author{Ying-Cheng Chen}
\affiliation{Institute of Atomic and Molecular Sciences, Academia Sinica, Taipei City, Taiwan}

\author{Hsiang-Hua Jen}
\affiliation{Institute of Atomic and Molecular Sciences, Academia Sinica, Taipei City, Taiwan}
\affiliation{Physics Division, National Center for Theoretical Sciences, Taipei City, Taiwan}

\author{Christoph Simon}
\affiliation{Institute for Quantum Science and Technology, and Department of Physics \& Astronomy, University of Calgary, 2500 University Drive NW, Calgary, Alberta T2N 1N4, Canada}


%
%
%
%
%
\begin{abstract}
Academia, governments, and industry around the world are on a quest to build long-distance quantum communication networks for a future quantum internet. 
Using air and fiber channels, quantum communication quickly faced the daunting challenge of exponential photon loss with distance.
Quantum repeaters were invented to solve the loss problem by probabilistically establishing entanglement over short distances and using quantum memories to synchronize the teleportation of such entanglement to long distances.
However, due to imperfections and  complexities of quantum memories, ground-based proof-of-concept repeater demonstrations have been restricted to metropolitan-scale distances.
In contrast, direct photon transmission from satellites through  empty space faces almost no exponential absorption loss and only quadratic beam divergence loss. 
A single satellite successfully distributed entanglement over more than 1,200 km.
It is becoming increasingly clear that quantum communication over large intercontinental distances (e.g. 4,000-20,000 km) will likely employ a satellite-based architecture. 
This could involve quantum memories and repeater protocols in satellites, or memory-less satellite-chains through which photons are simply reflected, or some combination thereof. 
Rapid advancements in the space launch and classical satellite communications industry provide a strong tailwind for satellite quantum communication, promising economical and easier deployment of quantum communication satellites.

\end{abstract}
%
\maketitle



\section{Introduction}
Quantum communication networks are emerging as an advanced communication technology while also offering new avenues for fundamental physics research~\cite{simon2017towards}. 
A quantum network enabling the transfer of quantum information anywhere on Earth, i.e. a quantum internet, would transform many current and future technologies. 
Among these technologies are quantum cryptography, quantum sensing and timekeeping, as well as distributed quantum computing ~\cite{bennet1984quantum, ekert1991quantum, bennett1992quantum, xu_secure_2020, wehner_quantum_2018, kimble_quantum_2008, gottesman_longer-baseline_2012, komar_quantum_2014, main2025distributed, aghaee2025scaling}. 
A form of quantum cryptography, quantum key distribution (QKD), provides information-theoretic security based on the laws of quantum physics ~\cite{bennet1984quantum, ekert1991quantum, bennett1992quantum}.
QKD relies on the distribution and measurement of both separable and entangled qubits and is the most near-term application of quantum communication in which significant progress has already been made~\cite{xu_secure_2020}. 
Distributed entanglement for quantum sensing includes improved precision of long baseline telescopes~\cite{gottesman_longer-baseline_2012} and entangled atomic clocks for improved timekeeping or navigation~\cite{komar_quantum_2014}.
In distributed quantum computing, multiple otherwise independent quantum computers are connected to achieve computation in a much larger Hilbert space and allow multi-party computational protocols~\cite{wehner_quantum_2018, liu2024road, main2025distributed, aghaee2025scaling}. 
Other cryptographic primitives in quantum networks include blind quantum computing~\cite{broadbent2009universal, arrighi2006blind, barz2012demonstration}, private database queries~\cite{giovannetti2008quantum, jakobi2011practical} and quantum secret sharing~\cite{hillery1999quantum}. 
Blind quantum computing allows users to run quantum computations on a remote server without revealing their inputs, computation or results.
Private database queries enables retrieval of specific information from a database without disclosing the query itself to the database owner.
Quantum secret sharing distributes a secret quantum state among multiple parties, requiring collaboration to reconstruct it, ensuring security and resilience to the loss of a subset of parties.
Fundamental physics research with quantum networks include tests of quantum mechanics like Bell inequality violations at large distances~\cite{yin_satellite-based_2017, hensen_loophole-free_2015} and simultaneous tests of quantum mechanics and general relativity~\cite{xu_satellite_2019, barzel2024entanglement, borregaard2024testing, rideout2012fundamental,mohageg2022deep}.

Information in quantum networks is carried by photonic quantum states.
The principal issue hindering the development of a quantum network is photon loss and, depending on the application, information errors.
Since quantum mechanics forbids qubit cloning (i.e. copying) to overcome loss ~\cite{wootters_single_1982, dieks_communication_1982, ghosh2004local}, several other methods have been investigated. 
One way is to merely generate qubits at a high rate and directly transmit them through fiber or air~\cite{ursin_entanglement-based_2007}.
However, this approach is practically constrained to short distances due to attenuation through these channels. 
Optical attenuation loss in fiber or air scales exponentially with distance. 
For example, optical fibers designed for low-loss classical communications have an attenuation coefficient as low as 0.16 dB/km (Corning Vascade). 
This corresponds to a loss of 100 dB or 10$^{-10}$ in 625 km (distance from Boston to Washington D.C.), corresponding to only 1 Hz transmission rate for a 10 GHz source~\cite{boaron_secure_2018}. 
The effect of attenuation and atmospheric turbulence, not to mention the curvature of the Earth, has limited the distance of experiments in ground-to-ground free-space optical links through air to only around 150 km~\cite{schmitt-manderbach_experimental_2007, xu_secure_2020}. 

Quantum repeaters were proposed~\cite{briegel_quantum_1998} to distribute entanglement over long and/or lossy channels by utilizing quantum memories. 
Once entanglement is distributed it is used directly or for teleporting quantum information.
In most repeater protocols, photons travel short distances to generate entanglement between near-separated quantum memories.
Entanglement between these memories is converted into entanglement between widely separated memories through entanglement swapping~\cite{sangouard_quantum_2011, duan_long-distance_2001, simon2017towards}. 
In this way, an improved loss scaling can be achieved over fiber and air, in principle.
However, other than entanglement sources repeaters also require high-performance quantum memories which, despite years of impressive progress, are still significant technological bottleneck for practical implementations.
There has been only a few repeater (sub-link) demonstrations that surpass the performance of direct transmission \cite{bhaskar2020experimental,lucamarini2018overcoming}. But rates are impractical and deployment in the field has been limited, although progress has been made in these directions~\cite{pu2021experimental, knaut2024entanglement, krutyanskiy2023telecom}.
Moreover, the simplest repeater protocols (sometimes referred to as "first-generation" repeaters~\cite{jiang2009quantum}) cannot reliably distribute quantum information beyond a certain distance (e.g., 2,000 km) due to the accumulation of errors and increasing experimental complexities of the repeater protocol \cite{simon2017towards}.
Other repeater concepts under development~\cite{zwerger_long-range_2018}, called second- and third-generation repeaters~\cite{jiang2009quantum, munro2010quantum} use quantum error correction to further extend distances ~\cite{muralidharan2016optimal}, including all-photonic repeaters~\cite{azuma2015all,munro2012quantum, muralidharan2014ultrafast} that aim to eliminate quantum memories entirely in a repeater. 
However, these protocols require extensive resources and frequent repeater stations~\cite{borregaard2020one, pant2017rate}.



Optical attenuation inside a medium (e.g., air or fiber) is not necessarily a limiting factor for global-scale quantum communication, if a communication medium with low attenuation loss is used.
However, it is challenging to create new kinds of optical fibers.
For instance, if long experimental fluoride glass fibers with theoretically predicted minimum loss of $6.5*10^{-3}$ dB/km~\cite{cozmuta2020breaking} could be created and produced in a scalable way, global-scale (i.e., 20,000 km) transmission would still have about 130 dB loss. 
Although such fibers can have low loss (in tens of dB) when used for quantum networks over thousands of kilometers, currently they still have higher losses than silica fibers.
An alternative for direct transmission is sending light through vacuum. 
This can be the vacuum of outer space accessed through satellite transmissions, which form the core subject of this review. Another recently proposed direction is "vacuum beam guide" channels through evacuated tubes, potentially buried into the earth~\cite{huang2024vacuum}, although they would potentially face enormous construction, political and financial challenges. 

Transmission of quantum information using satellites has been remarkably successful and research is progressing rapidly~\cite{yin_satellite-based_2017, yin_entanglement-based_2020, ren_ground--satellite_2017, yin_entanglement-based_2020, liao_satellite--ground_2017, liao_satellite-relayed_2018}.
Photonic qubits mostly travel through empty space without any exponential attenuation loss, due to lack of a medium.
The only attenuation loss associated with satellite transmission is the short distance ($\sim$20 km) light needs to travel into and out of the earth's atmosphere for connecting ground stations. 
For vertical transmission (i.e., at zenith) at 1550 nm telecommunication wavelength atmospheric attenuation loss is around 10$\%$ (or 0.5 dB) ~\cite{gagliardi1995optical}.
Photons from satellites instead face diffraction loss or beam divergence loss which increases only quadratically with distance. 
Due to this major advantage, the Micius satellite launched by China in 2016 achieved a groundbreaking entanglement distribution distance of 1,200 km~\cite{yin_satellite-based_2017} and performed entanglement-based QKD over 1,120 kms~\cite{yin_entanglement-based_2020}, going much further beyond optical fibers. 
However, due to Earth's curvature single satellites in low-earth orbit (LEO, 200-2000 km altitude) like Micius have a visibility of the Earth only up to a maximum of a few thousand km. 
In addition, highly oblique (i.e. near-tangential) transmission through the atmosphere experiences higher absorption losses, further limiting range. 
To reach longer distances one may employ satellites in higher-orbits like geostationary orbits (at 36,000 km). However, such high orbits will cause large diffraction loss and consequently achieve low rates.
This method is also associated with high communication latency time with the satellite.
To reach intercontinental and global distances (say, 4,000 to 20,000 kms), several quantum repeater proposals combining satellites and quantum memories have been put forward. 
This includes architectures with quantum memories in ground stations~\cite{boone_entanglement_2015} or on-board satellites~\cite{gundogan_proposal_2021, liorni2021quantum}, including schemes where satellites move to transfer stored photons~\cite{simon2017towards, gundogan2024time}.
Recently, a simpler architecture was proposed to achieve global scale entanglement distribution by just reflection, without requiring quantum memories or repeaters~\cite{goswami2023satellite}. 
This optical relay approach utilizes a synchronously moving satellite chain where light is refocused at each satellite in the chain and transmitted to the next.
Such a relay confines beam divergence indefinitely akin to a set of lenses on an optical bench.
Such proposals, combined with rapidly advancing satellite experiments, diversify the possibly mission architectures for deploying satellite-based quantum networks.

Satellite-based quantum communications are poised to leverage the rapid progress in the space industry, including low launch costs offered by reusable rocket technology~\cite{elvis2023accelerating, palmer2021spacex, heldmann2022mission, jozivc2020fuel, lao2024study, lee2020heavy} and an exponentially growing constellation of satellites~\cite{brashears2024achieving,chaudhry2021laser, henri2020oneweb, osoro2021techno}. 
This offers lower cost and easier launches for quantum satellite networks, and broader implementations of quantum technologies in space.

In this review, we first overview quantum repeaters and their challenges.
This motivates satellite-based quantum communication. We then introduce satellites and discuss their advantages and limitations.
Low-earth satellite experiments using Micius are discussed. To achieve longer distances, several protocols are discussed including those that use satellites and quantum memories.
Finally, the satellite relay is introduced, discussing its benefits and chanllenges, including comparison to the to ground-based relay proposal called vacuum beam guides.
Before concluding, we explore the potential of a combined relay-repeater network.
Parts of this review are inspired by the quantum network discussions in Ref. ~\cite{goswami2021photonic}.

\section{Quantum Repeaters}

\begin{figure}[htbp]
    \centering
    \includegraphics[width=0.75\linewidth]{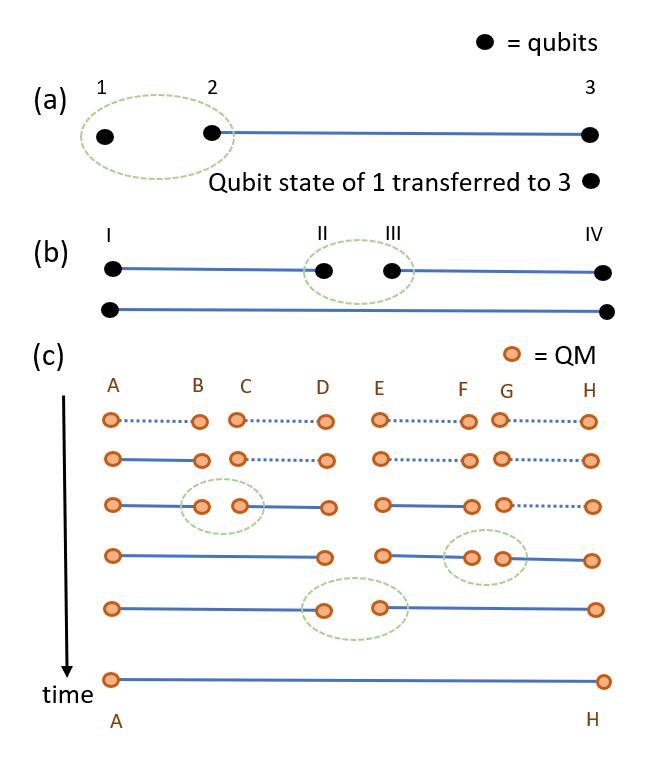}
    \caption{(a) Teleportation - Initially distant qubits 2 and 3 are entangled (shown by solid blue line), while qubit 1 is separate with its own quantum state but physically close qubit 2. In the next step, the qubit state of 1 is teleported to the distant qubit 3 by doing a joint operation on qubits 1 and 2. This operation (shown by green dashed circle) consists of an entangling gate followed by measurement. For photonics qubits this can be Bell-state measurement, while for atomic qubits it is a two-qubit entangling gate followed by measurement. Further classical communication of the measurement results and single qubits gates on qubit 3 based on the measurement results is needed to complete the teleportation process. This is not shown in the figure for simplicity and mentioned in detail in text. (b) Teleporting an entangled qubit (say II here) to a distant location (say, at the place of IV) forms the basis of a quantum repeater. This enables entangled distribution over longer distances. (c) A schematic diagram showing the working principle of quantum repeaters. Initially, eight quantum memories (A-H) are not connected to each other. Eventually, entanglement between A and H was established though successive tries of entanglement generation, storing and eventually performing the entanglement swapping operation (See text for details). The figure is reproduced from~\cite{goswami2021photonic} with permission.
}
    \label{QN_chapter_repeater_expln}
\end{figure}

The goal of a quantum network is to distribute quantum information between different locations which can be separated by long and/or lossy channels. 
Quantum repeaters are designed to solve the photon loss problem by cleverly synchronizing the distribution and teleportation of entanglement over a channel using quantum memories. 
Then quantum information can be transferred by teleportation using the entanglement established between the end points of the channel.
The concept of a quantum repeater is shown in Fig. \ref{QN_chapter_repeater_expln}. 
The basic scheme of teleportation, upon which quantum repeaters rely on, is shown in Fig. \ref{QN_chapter_repeater_expln}(a).
The state of one qubit (1) is teleported to a distant qubit (3) entangled with another qubit (2) when a joint measurement is performed on qubits 1 and 2.
This requires pre-established entanglement between distant qubits 2 and 3. 
The joint measurement consists of an entangling gate followed by measurement (shown by a green dashed ellipse) which can be implemented using linear optics or matter qubits. 
Afterwards, the measurement result is classically transmitted to properly interpret qubit 3 and complete the teleportation. 

In a quantum repeater, entanglement is distributed over several short segments of the channel, and a chain of successive teleportation events between one of the qubits from each entangled state leads to entanglement being established between the ends of the channel.
This is shown in Fig. \ref{QN_chapter_repeater_expln}(b) for dividing the channel into two segments; the state of II is teleported to IV.
Such teleportation of entanglement is also referred to as entanglement swapping.
However, one can only perform teleportation if the joint measurement (between II and III in Fig. \ref{QN_chapter_repeater_expln}(b)) succeeds, that is, the entangled qubits must not be lost during their distribution over the short segments.
The probability of this occurring directly (without storage in memories) is equivalent to a single photon successfully traversing the entire channel (from I to IV in Fig. \ref{QN_chapter_repeater_expln}(b)), rendering the teleportation approach no more effective than direct transmission.

Quantum memories address this challenge by storing entangled qubits until entanglement is successfully generated across neighbouring segments in a quantum repeater. 
This synchronizes the arrival of the entangled qubits for the required joint measurements for teleportation/swapping, and heralding thereof, which extends the entanglement over longer distances. 
Figure \ref{QN_chapter_repeater_expln}(c) illustrates this process. 
Initially, eight separated quantum memories (A-H) attempt to establish entanglement between neighboring pairs (e.g., A-B, C-D). 
Successful attempts are stored while failed ones (e.g., C-D in first attempt) are repeated. 
Once adjacent pairs are entangled (e.g., A-B and C-D), joint entangling measurements are performed on nearest neighbor qubits (e.g. B and C) resulting in entanglement being established over longer distances (e.g. A and D). 
This process continues, ultimately entangling the end points (A and H). 
Although simplified, this example demonstrates the fundamental principle of quantum repeaters.

In our example, the longest time certain quantum memories  (e.g., memory A) need to store a qubit is from the time entanglement is first created in the memory, until entanglement distribution in the whole link is finished. 
Hence, quantum memories must have long storage times. 
Depending on the architecture, the minimum storage time ranges from seconds to a millisecond, and is fundamentally bounded by the light travel time in the short segments. 
The length of each segment must also be long enough for practical use and to avoid the compounding effects of imperfections at the storage and swapping steps.
Quantum memories must also be efficient and low noise, accommodate a reasonable entanglement distribution rate, ideally store a large number of entangled modes, and must be compatible with the entangling operation step.
These properties are crucial, as only a few entangled pairs are successfully stored out of the many many photons sent even when near-ideal quantum memories are used~\cite{simon2017towards, sangouard_quantum_2011}. 
The entanglement generation rate can also be linearly increased using multiplexing of memories and channels, e.g. through use of additional spatial, temporal or spectral modes.
The strict requirements on the quantum memories represents one of the most significant barriers towards successful implementation of a quantum repeater. 
Milestone properties have been individually approached in separate quantum memories~\cite{lei2023quantum}, but not all together in a single device. 
For instance, ultracold atomic memories and solid state memories have both achieved over 100 ms storage time~\cite{wang2021cavity, korber2018decoherence, ma2021one} and separately above 90$\%$ efficiency in ultracold atoms~\cite{hsiao2018highly, wang2019efficient}. 
On the other hand rare-earth ion doped atomic frequency comb quantum memories have achieved large multiplexing capacity by storing as many as 1650 separate single photon modes~\cite{wei2024quantum, bonarota2011highly, sinclair2014spectral}. 
Silicon vacancy centers in diamond have demonstrated a repeater that improves over direct transmission, but these memories are low rate and currently lack multiplexing capabilities~\cite{bhaskar2020experimental}.
In later sections in this review, we will discuss quantum memories in the context of those that store and retrieve externally provided photons, although different types of memories, e.g. those that produce probabilistic or deterministic spin-photon entanglement, may be used in principle with suitable protocol modification.

The fundamental structure of the quantum repeater described above was proposed by Briegel et. al.~\cite{briegel_quantum_1998}. 
One of the early proposals to implement such a quantum repeater using atomic ensembles and heralded entanglement generation was proposed in 2001 by Duan, Lukin, Cirac and Zoller~\cite{duan_long-distance_2001}, widely known as the DLCZ protocol. 
Although seminal, the limitations of DLCZ include low rates to avoid creation of multiphoton-pair events, inefficient heralding, lack of multiplexing and operation at visible wavelengths, which is incompatible with long-distance fiber transmission, in addition to the challenge of achieving high-performance quantum memory~\cite{radnaev2010quantum, wein_efficiency_2016}.
To solve these and other problems, many different repeater schemes have been designed over the years~\cite{jiang_fast_2007, chen_fault-tolerant_2007, simon_quantum_2007, collins_multiplexed_2007, azuma2023quantum}. 
For example, one of these uses a non-degenerate entangled photon pair source~\cite{fekete2013ultranarrow} and multiplexed memories. 
One of the photons emitted by the entanglement source is at telecom wavelengths to be sent to long distances, whereas its pair photon can be at a suitable wavelength for storage in quantum memory~\cite{afzelius2009multimode, afzelius2010efficient}. 
The quantum memories and entanglement sources used in many quantum repeater proposals are based on (bulk) ensemble systems, such as cold atoms and nonlinear crystals, respectively.
Repeaters using individual quantum systems (i.e. single atoms, ions, color centers in diamond, quantum dots etc.) for entanglement generation and storage has also experienced a lot of attention and progress~\cite{barrett_efficient_2005, hensen_loophole-free_2015, pompili_realization_2021, delteil2016generation, stockill2017phase, krutyanskiy2023entanglement, rosenfeld2017event, bhaskar2020experimental, knaut2024entanglement}. 
The principal scheme is generally similar: creating and or storing excitations in remote stations, and upon readout, projecting them into an entangled state by erasing the which-way information that the photons possess~\cite{hensen_loophole-free_2015}. 
The primary advantage of single systems over ensembles is suppression of multi-photon events and noise, deterministic gates, and built-in qubits for error correction \cite{bhaskar2020experimental, duan_long-distance_2001, borregaard2015long}.
However, such systems can be hard to create at scale, and hence have been challenging to interface and multiplex, in addition to requiring carefully-constructed cavities to enhance light-matter coupling.

Research on quantum repeaters and physical systems to construct them is ongoing.
In the meantime, proof-of-concept quantum repeater has been implemented in lab settings and also recently in real-world scenarios over up to 50 km in several metropolitan areas~\cite{ pu2021experimental, bhaskar2020experimental, langenfeld2021quantum, yu2020entanglement, liu2024creation, knaut2024entanglement, stolk2024metropolitan,  krutyanskiy2023telecom, castelvecchi2024quantum, liu2024road}. 
However, quantum repeaters are just beginning to surpass the rate-loss bound \cite{pirandola2021satellite} for direct transmission experiments that are outside of the lab \cite{bhaskar2020experimental,lucamarini2018overcoming,knaut2024entanglement}. 
A rate-loss bound is often used to distinguish direct transmission from a repeater independent of its specific components or protocol.
Current demonstrations surpassing this bound have been very low-rate in that the demonstrations allowed non-zero transmission of qubits over a channel of exceptionally high loss, e.g. \cite{bhaskar2020experimental}.
The bound is generally defined with respect to a single (non-multiplexed) channel or mode, and does not take into account requirements to implement any specific protocol, e.g. QKD or distributed quantum computing.

Due to the complexities in their design and implementations, existing proof-of-concept repeater demonstrations generally fall into the first-generation quantum repeater category. 
There are other proposals~\cite{muralidharan2016optimal} that employ quantum error correction in quantum repeaters to actively combat noise, called second-generation repeaters~\cite{jiang2009quantum, munro2010quantum}. 
Error correction enables distribution of higher-quality entanglement and allows reaching longer distances, in principle. 
There also exist proposals to eliminate the need for quantum memories altogether, called all-photonic repeaters~\cite{azuma2015all}, by sending a large encoded quantum states like graph states which will be error-corrected for both channel loss and operation losses~\cite{munro2012quantum, muralidharan2014ultrafast}. 
However, both of these alternatives have extremely high resource requirements and are only very early on in their development. 
For example, they require many qubits as resources and very frequent repeater stations, e.g. every 1-2 kms~\cite{borregaard2020one}. 
Resource estimation shows that one needs 10$^6$ single photon sources per node, even when repeater stations are separated by just 1.5 km~\cite{pant2017rate}. 
Due to the aforementioned challenges with repeater architectures, direct transmission of qubits from orbiting satellites have been investigated to reach large distances with high rates. 

\section{Satellites}

Quantum communication satellite missions have achieved entanglement distribution over 1200 km~\cite{yin_satellite-based_2017} between two locations on earth, which is impossible currently with any fibre-based architecture. 
This has been possible as loss in satellite transmission is principally due to diffraction, which varies quadratically with distance as compared to the exponential attenuation loss in optical fibers or air transmission. 
Diffraction loss occurs due to beam divergence. 
This increases with communication distance and also depends on the transmitting and receiving telescope diameters. 
Large ground telescopes (meter diameters) are commonly used in space-to-ground quantum communication demonstrations. 
However, satellite telescopes are restricted to smaller sizes (in tens of cm) as large telescopes are expensive to deploy in orbit. 
Beam pointing error also an important contributor to channel loss~\cite{gagliardi2012satellite}.
A LEO satellite circles the earth almost every 90 minutes at 8 km/s speed. 
Hence, employing sophisticated satellite tracking technology is one of the most important parts of a satellite-based quantum communication mission~\cite{bourgoin_comprehensive_2013}.


Atmospheric transmission loss, due to absorption, scattering and turbulence, is another significant factor. 
Absorption loss depends on wavelength of light and angle of incidence through the atmosphere~\cite{bourgoin_comprehensive_2013}. 
Light to and from satellites that is incident at a very low elevation angle incurs dramatically increased attenuation as it traverses a long path through the atmosphere~\cite{kasten1989revised}.
Turbulence generates random fluctuations in the refractive index of the air which results in beam spreading, beam wondering and even beam fragmentation, causing loss. 
Turbulence is most pronounced close to the surface of the earth as atmosphere is thickest closest to the Earth. After 20 km, the atmosphere itself is too thin to be considered. 
Generally, turbulence impacts uplink transmission to a greater extent than downlink transmission, because in uplink the turbulence induced beam, after emanating from the atmosphere,  has to travel for hundreds of kms to reach the satellite.


Adaptive optics systems can help compensate for turbulence by first surveying the atmosphere using a downlink reference light from the satellite and then introducing beam tilt and wavefront corrections to the outgoing up-link beam using segmented-mirror ground telescopes. 
Atmospheric turbulence effects occur on time scales of the order of 10-100 ms (but can be much shorter in some conditions). 
Laser pulses, and qubit wavepacket durations, are generally much shorter than that timescale enabling them to properly survey and compensate for the refractive index variations.
Such compensation does not work very well for up-links as the downlink reference beam from satellite and the uplink beam containing the qubit cannot be in the same spatial mode due to the 8 km/s satellite motion.
Better results can be obtained using laser guide stars (LGS) which are artificial stars created by exciting sodium in the mesosphere at around 90 km elevation using a powerful laser that follows satellite movement~\cite{pugh_adaptive_2020}. 
The light from this artificial moving star travels through the turbulent atmosphere to be detected on earth for compensation.
These different adaptive optics methods, developed in classical optics (especially for astrophysical observations), provide partial solutions to the high channel loss issue in the up-link due to turbulence.
In quantum communication these techniques are inherited while making sure the quantum channel are not corrupted (i.e, qubit decoherence) due to the turbulence effect. 
Polarization~\cite{ren_ground--satellite_2017} and time-bin~\cite{villasenor2021enhanced} qubits can be used for quantum communication through the uplink channel.


LEO satellites - either in uplink or downlink - face the challenge of limited flyby times.  
These satellites typically move at speeds around 8 km/s.  
For a satellite in a 500 km-distance orbit, assuming tracking over a 1000 km distance, the flyby time is approximately 2 minutes. 
Tracking at higher elevation angles, where atmospheric loss is greater, can slightly extend this time, but it remains limited still to a few minutes for LEO satellites. 
While LEO satellites have shorter flyby times per ground station, their movement allows them to rapidly service different locations as they traverse their orbit. 
As the Earth also rotates below the rotating LEO satellite, a properly positioned LEO satellite can fly over every point in Earth within a specific interval. 
For example, a satellite in polar orbit will pass over and be able to transmit to any specific point on earth at least twice each day, in clear weather. 
Quantum communication with LEO satellites can take advantage of this for addressing several sites across the world at regular fixed intervals.

The signal-to-noise ratio (SNR) plays a critical role in quantum communication. 
Noise shortens communication distances and restricts or prohibits quantum network functionalities.  
During daytime, reflected sunlight becomes the dominant noise source.  
Consequently, daytime operation of satellite QKD remains a challenge, as demonstrated by the Micius satellite, in which links are predominantly established at night. 
At night, the dominant noise source shifts to light pollution from cities. 
Therefore, ground stations located far from urban centers, where light pollution is minimized, are preferable for achieving optimal SNR. 
High SNR can be achieved by reducing background noise and/or increasing the signal strength, through reduced transmission loss, improved filtering, or enabling large multiplexing capacity for the signal \cite{raymer2020time, liao2017long}. 
For example, in a recent paper~\cite{abasifard2024ideal} a novel idea was proposed to enable daytime operations. The large background noise from sunlight can be reduced if the quantum signal employs particular wavelengths that are absent in the solar spectrum or have a very reduced intensity (e.g., the dark absorption lines or Fraunhofer lines)~\cite{abasifard2024ideal}. Adaptive optics systems has been used to reduce daytime noise in free-space-optical QKD channels too~\cite{gruneisen2021adaptive}.


\section{Satellite transmission experiments}


Prior to the Micius satellite launch there were several feasibility studies to ascertain parameters like total loss or background noise in downlink or uplink transmission~\cite{bonato_feasibility_2009, bourgoin_comprehensive_2013}. 
Initial experiments were also conducted using optical sources or retroreflectors present in existing satellites. 
These experiments investigated different parameters, including conclusively proving that a polarization qubit would not decohere due to turbulence in atmospheric transmission~\cite{toyoshima_polarization_2009,bedington_progress_2017}.
In addition to these test experiments, retroreflectors can also achieve downlink qubit transmission by working as polarization converters to uplink weak laser pulses sent from ground to the satellite~\cite{bedington_progress_2017}.

\begin{figure}[htbp]
    \centering
    \includegraphics[width=0.95\linewidth]{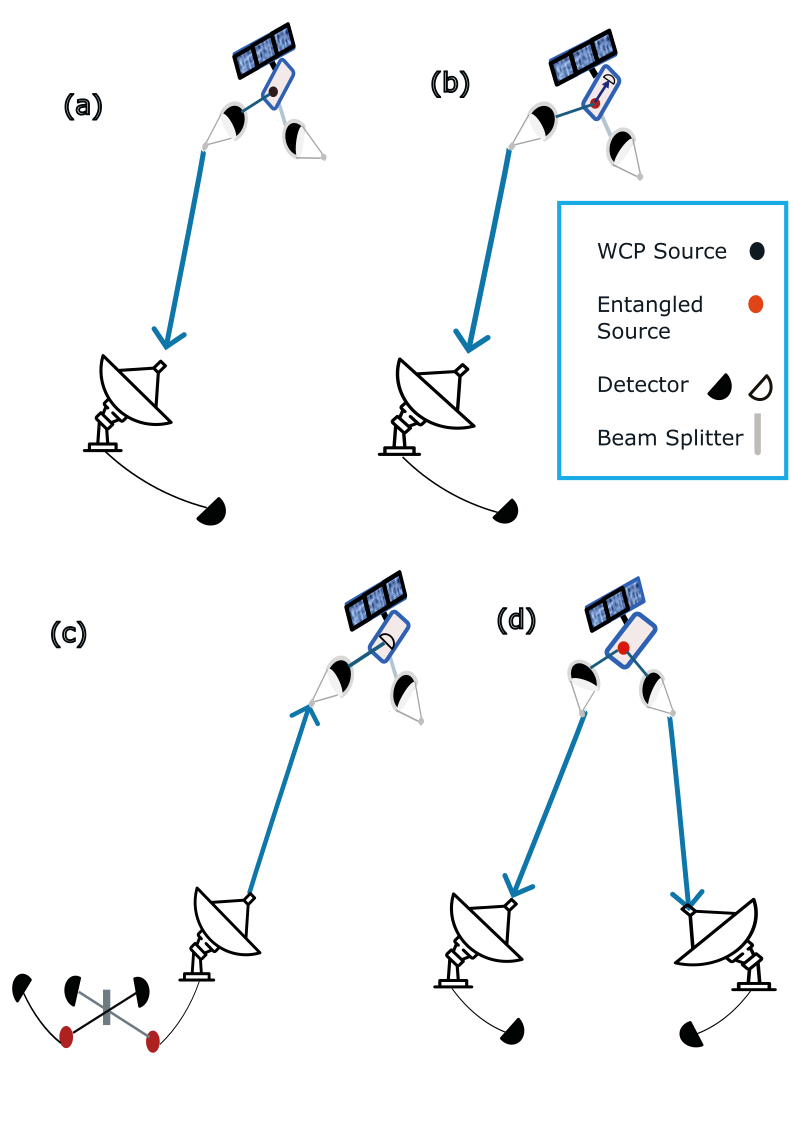}
    \caption{Different experiments performed with the Micius satellite~\cite{yin_satellite--ground_2017, yin_satellite-based_2017, ren_ground--satellite_2017,liao_satellite--ground_2017}, as described in~\cite{goswami2021photonic}. (a) Downlink QKD - Using a weak coherent pulse (WCP) source (black circle) photons are sent downlink to ground station to perform decoy-state QKD~\cite{liao_satellite--ground_2017}. (b) Entanglement based QKD - An entangled pair source (red circle) aboard Micius is used to perform QKD between the satellite and the ground station. (c) Uplink teleportation - Entangled photon pair source (red circle) in ground station is used to teleport a qubit to the satellite. The unknown qubit, to be teleported, also comes from another entangled pair for technical reasons. See text for details~\cite{ren_ground--satellite_2017}. (d) Entanglement distribution in downlink- Entangled photon pairs are distributed between two ground stations, separated by a record 1203 km on earth, by double downlink transmission~\cite{yin_satellite-based_2017}.
}
    \label{QN_review_fig2}
\end{figure}

In 2017, in culmination to all the previous work, the Micius satellite was launched by China which achieved several milestones: entanglement distribution over 1200 km~\cite{yin_satellite-based_2017}, quantum teleportation to the satellite over 1400 km~\cite{ren_ground--satellite_2017}, and several successful QKD demonstrations~\cite{yin_satellite--ground_2017, liao_satellite--ground_2017, liao_satellite-relayed_2018, yin_entanglement-based_2020, lu2022micius}. 
The different experiments performed by Micius satellite are depicted in Fig. \ref{QN_review_fig2}. 
Micius was equipped with two telescopes (with diameters 30 cm and 18 cm), PPKTP (Periodically Poled Potassium Titanyl Phosphate)-based SPDC (Spontaneous Parametric Down-Conversion) entangled photon source, a weak coherent pulse (WCP) source to carry out QKD schemes and single photon detectors. 
It also had more lasers, alignment optics and detectors to be used for satellite tracking (acquiring pointing and tracking (APT) system) and polarization correction. 
Micius performed QKD in the down-link scenario using a WCP source through the BB84 protocol~\cite{liao_satellite--ground_2017} and another down-link QKD protocol using onboard entanglement source~\cite{yin_satellite--ground_2017}, as shown in Fig. \ref{QN_review_fig2}(a)-(b). 
Micius also performed an up-link transmission experiment. 
As depicted in Fig. \ref{QN_review_fig2}(c), one qubit from an entangled pair was up-linked to Micius and an entangled photonic qubit was teleported/swapped from ground to the satellite over distances ranging between 500-1400 km~\cite{ren_ground--satellite_2017}. 
Both entangled photon pairs were generated from the same pump laser for phase control and to ensure complete mode overlap of the photons on a beam splitter to implement the entangling gate and measurement needed for teleportation.
Micius also distributed entanglement between two ground stations in China separated by 1203 km~\cite{yin_satellite-based_2017}, as schematized in Fig. \ref{QN_review_fig2}(d). 
Notably, they performed a Bell inequality test on the entangled pair, a measurement of non-locality and hence non-classicality, which increased the Bell test violation distance by almost an order of magnitude. 
Entanglement-based QKD was also achieved over 1,120 km in 2020 using this configuration~\cite{yin_entanglement-based_2020}. 
Several so-called trusted-node QKD experiments were performed using Micius.
In this scenario, the satellite acts as a trusted party and generates a unique key to be shared with each of the ground stations~\cite{liao_satellite-relayed_2018, chen_integrated_2021}. 
In one such experiment, a quantum-encrypted video call was conducted between Austria and China~\cite{liao_satellite-relayed_2018}. 
More recently, trusted-node QKD has been performed over 4600 km between two cities in China, with trusted nodes in both ground and satellite~\cite{chen_integrated_2021}. 
Launching a whole fleet of satellites to perform Global trusted node QKD has been considered~\cite{noauthor_china_2024}. 
Indeed, trusted nodes cannot distribute entanglement and hence cannot enable other functionalities of quantum internet like distributed quantum computing, sensing or teleportation. 
To accomplish them worldwide along with completely secure QKD, global-scale entanglement distribution is needed.
Trusted-node QKD using satellites have been viewed as more secure than using ground stations as trusted nodes because satellite nodes are remote in space. 
This may change with advances in space technology if access to space drastically improves.



After the success of the Micius satellite, several other satellite projects have been announced, initiated or implemented. 
A smaller micro-satellite called Jinan-1 was launched in 2021 to perform QKD~\cite{li2024microsatellite}. 
China is also considering launching more satellites soon, including higher-orbit satellites for more coverage~\cite{noauthor_china_2024}. 
A cubesat called SpooQy-1, developed by the National University of Singapore, tested an entangled source in orbit in 2019~\cite{villar2020entanglement}. 
Several countries are also working towards their own quantum satellite mission, including Canada (QEYSSat satellite)~\cite{jennewein2023qeyssat, jennewein2024qeyssat}, the European Union (Eagle-1 small satellite made with SES)~\cite{rivera2024building}, the United States (SEAQUE - Space Entanglement and Annealing QUantum Experiment)~\cite{ortiz2024seaque}, and India (quantum satellite announced by ISRO)~\cite{singh2023india}. 
Private companies are also actively working towards developing quantum communications using satellites including QuantumCTek in China (backed by the Micius satellite team), Boeing's quantum division, Spectral in Singapore etc. 

However, single LEO satellites cannot establish entanglement beyond $\sim$ 2000-3000 km due to the curvature of earth.
Higher-orbit satellites are one way to achieve this.
Satellites in geostationary (GEO) orbit (at 36,000 km elevation) or even mid-earth orbits (MEO - orbits between LEO and GEO) can establish entanglement between two farther away places in earth.
Diffraction loss will be very high though for high orbits compared to LEO satellites. 
However, while LEO satellites have limited flyby times, higher orbits offer longer flyby times, with geostationary (GEO) satellites being effectively stationary relative to a ground station. 
This continuous availability increases the communication rate. 
However, even for this continuous coverage of two specific ground stations, GEO satellites are expected to have a very small rate of entanglement distribution ($\sim$ 1 Hz)~\cite{boone_entanglement_2015}. Experiments have been performed using retroreflectors in higher Earth orbits to model the high- orbit transmission rates~\cite{vallone_experimental_2015, calderaro2018towards}. 
Even geostationary satellites cannot send photons to global distances (10,000-20,000 km, not even with 1 Hz rate), due to Earth's curvature and grazing incidence~\cite{boone_entanglement_2015} .


\section{Memory-Satellite protocols}
Satellite schemes combined with quantum memories offer a way to achieve ultra-long distance (more than 10,000 km) distribution of entanglement.
These could involve quantum repeaters over satellite chains or memories in single satellites.


\subsection{Repeater protocol with satellites}


\begin{figure}[b]
    \centering
         \begin{subfigure}
         \centering        \includegraphics[width=0.5\textwidth]{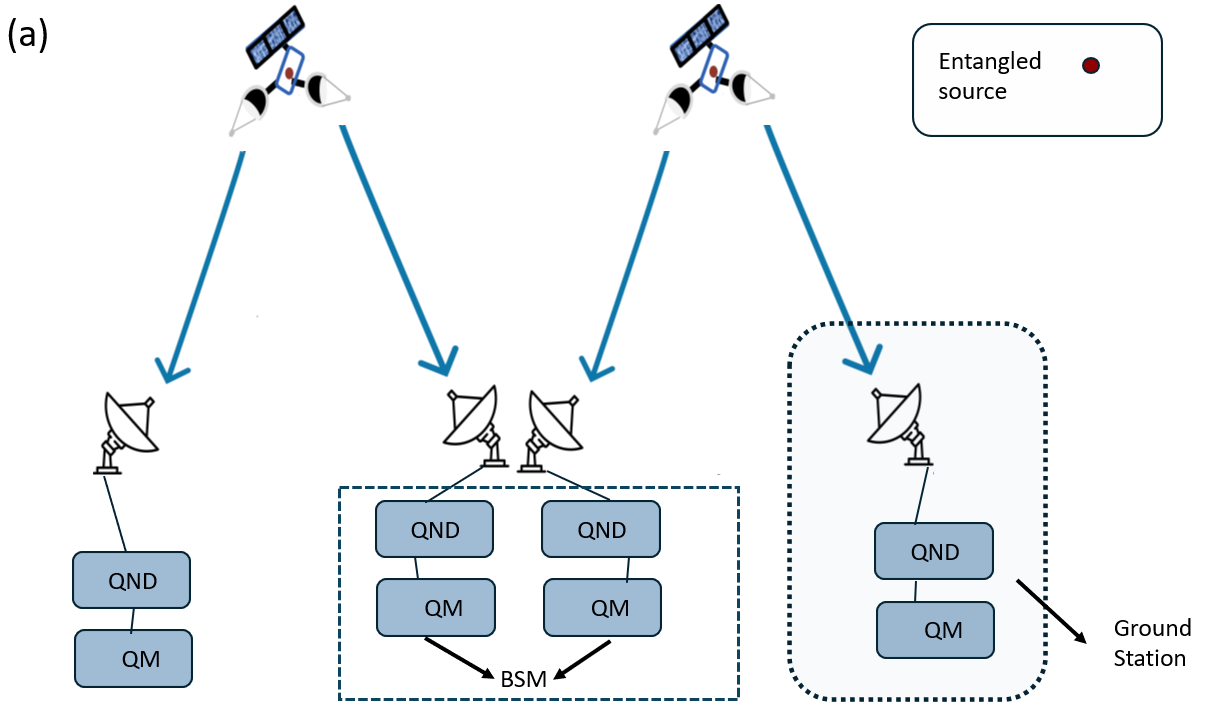}     \end{subfigure}
         \begin{subfigure}
         \centering        \includegraphics[width=0.5\textwidth]{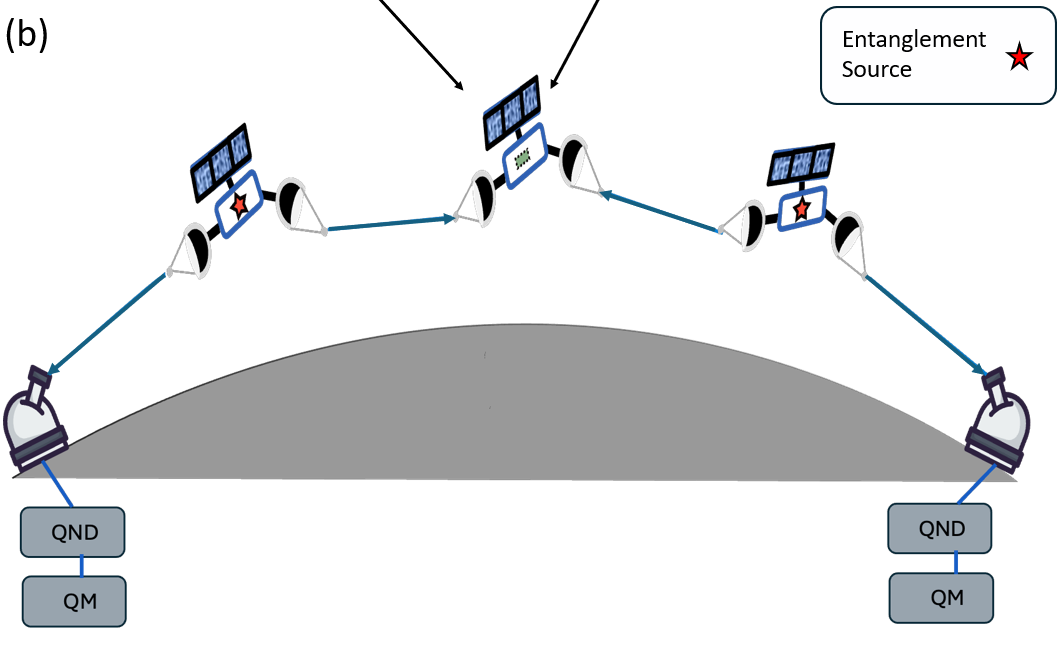}     \end{subfigure}         
    \caption{Two repeater protocols over LEO satellites to distribute entanglement over global distances using quantum memories (QM) and quantum non-demolition (QND) detectors in ground stations~\cite{boone_entanglement_2015}. In (a) the QM and QND detectors are kept in ground stations for operation advantages while in (b) the QM and QND detectors in the intermediate links (dashed box in (a)) are kept in the satellite to avoid extra intermediate ground link connections \cite{gundogan_proposal_2021, liorni2021quantum}.
    }
    \label{Fig_sat_repeater}
\end{figure}

The first such protocol was given in~\cite{boone_entanglement_2015}  
(shown in Fig. \ref{Fig_sat_repeater}(a)). 
Quantum non-demolition (QND) detectors are used for this protocol along with quantum memories. 
QND detectors detect the presence of a photon without disturbing its quantum state~\cite{braginsky_quantum_1980, grangier_quantum_1998, goswami_theory_2018, o2016nondestructive, reiserer2013nondestructive}. 
Both QND detectors and memories are kept on the ground for the sake of easy operation and maintenance while satellites will direct entangled photon pairs to the ground stations. 
The successfully transmitted entangled pairs are detected by QND detectors and subsequently stored in memories. 
Entanglement swapping between qubits stored in quantum memories is then performed, as shown in Fig. \ref{Fig_sat_repeater}(a), using a projective measurement onto a Bell state. 
In~\cite{boone_entanglement_2015} an equatorial satellite constellation was proposed. 
However, this method will work more generally for a constellation of satellites to distribute entanglement between any two points on earth. 


Numerical simulations consider 4 to 8 links between the equatorial satellite chain, with the possibility of different orbital heights (500-1500 km). Simulation parameters include 50 cm satellite transmitter diameters, 1 m ground telescope diameter, and 580 nm photon wavelength for the satellite-repeater transmission (motivated by Eu-doped yttrium orthosilicate memory storage wavelength). 
Efficiencies for sources, memory read and write operations, and single-photon detection is assumed of 0.9.
Efficiency for QND detectors is assumed to be 0.32. Also, repeater rates are explored in variation of these efficiencies.
The numerical results (shown in Fig. 2 of \cite{boone_entanglement_2015}) demonstrate entanglement distribution rates per day as a function of total ground distance for direct transmission from high orbits and satellite-repeaters. The rates for geostationary satellite direct transmission and repeaters are close around 4,000 km at about $10^4-10^5$ qubits/day ($\sim$ 1Hz). However, for longer distances (above 10,000 km) geostationary satellite rates fall off drastically, while satellite repeaters at 1000-1500 km orbits with 8 links still provides rates around 10$^3$ qubits per day at 20,000 km. The difference in rates also stems from considering a lower memory bandwidth for repeaters. Correspondingly, repeater calculations use a 10 MHz pair source repetition rate, while direct transmission uses 1 GHz. So, with higher memory bandwidth repeater rates can be increased by two orders of magnitude. Moreover, frequency multiplexing has not been considered in \cite{boone_entanglement_2015} either for direct transmission or for satellite-repeaters, which can increase rates further.

Quantum memories and QND detectors  can alternatively be in satellites~\cite{gundogan_proposal_2021, liorni2021quantum} (shown in Fig. \ref{Fig_sat_repeater}(b)). 
Given recent atomic and optical physics experiments in space~\cite{liu_-orbit_2018, aveline_observation_2020, gundougan2021topical} and the development of transportable, standalone quantum memories~\cite{jutisz2024standalone}, this possibility  is becoming more plausible. 
QND and QM placed in satellite would evade the extra downlink and atmospheric loss multiple links simultaneously for successful entanglement transfer over a long distance, as well as potentially reduce latency of the individual links.
Space-based memories would entirely bypass the need for intermediate  ground-space links, removing any dependence on weather conditions, except for the end links. This also obviates the need to place these terrestrial repeater stations in possibly inaccessible or inconveniently located areas (such as in the middle of an ocean). Coordination of dynamics of the different link geometries is also simplified. 


Simulations in \cite{gundogan_proposal_2021} compare the space-based memory proposal with the earlier ground-based memory proposal \cite{boone_entanglement_2015}. The performance metric used is the total time to distribute entangled pairs. A total of 8 links corresponding to repeater nesting level of 3 is assumed~\cite{duan_long-distance_2001}. Other parameters include  beam divergence of 4 $\mu$rad, total memory efficiency and QND detector efficiency of 0.9 while a source rate of 20 MHz. Numerical results presented in Fig. 2 of \cite{gundogan_proposal_2021} demonstrates three-orders of magnitude higher entanglement distribution rate using space based memory-QND system than ground based memory-QND. This also translates to significantly lower memory storage time requirements. Variation of repeater rates with different beam divergence values (ranging from 1 to 10 $\mu$rad), and memory efficiency (ranging from 0.5 to 1) is shown.
More recent studies has shown that memory optimization can increase rates even further~\cite{wallnofer2022simulating}.
A recent proposal involving single atom systems - both as single photon sources and memories - showed their architecture lowers the required multiplexed capacity~\cite{tubio2024satellite}.  
Note that these protocols could be applied to conventional ground-based repeater schemes and there have been other conventional repeater proposals that use QND \cite{azuma2012quantum}.

\begin{figure*}[htbp]
    \centering
             \begin{subfigure}
         \centering        \includegraphics[width=0.4\textwidth]{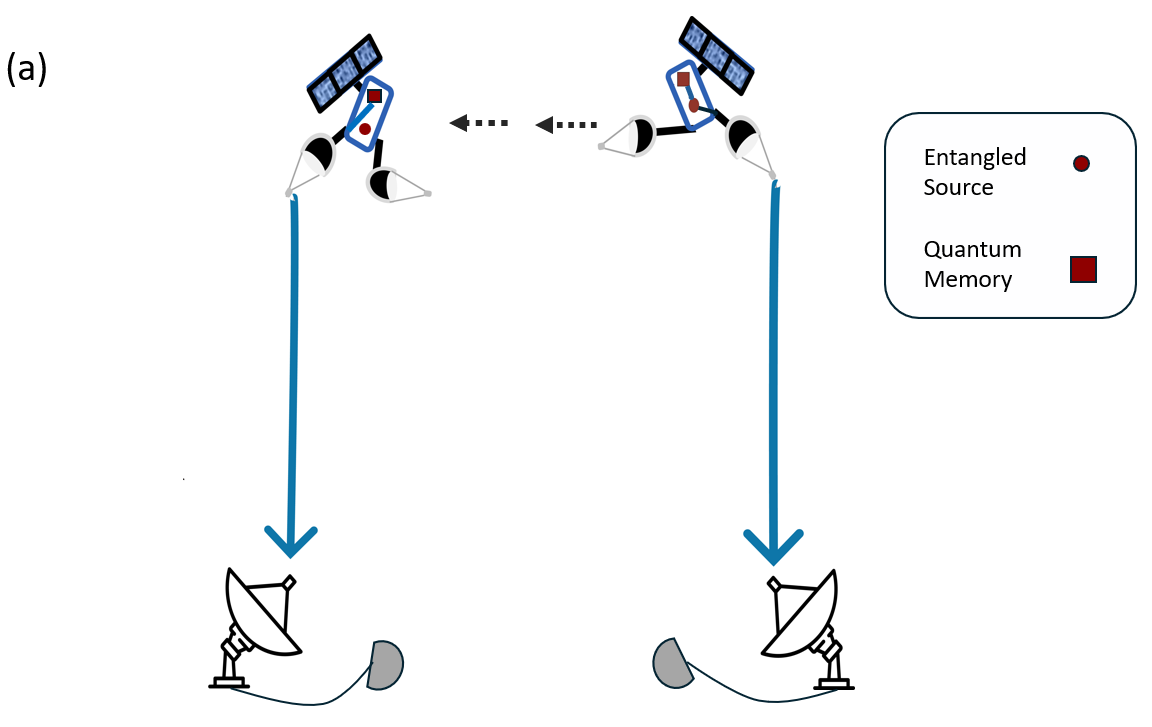}     \end{subfigure}
         \begin{subfigure}
         \centering        \includegraphics[width=0.55\textwidth]{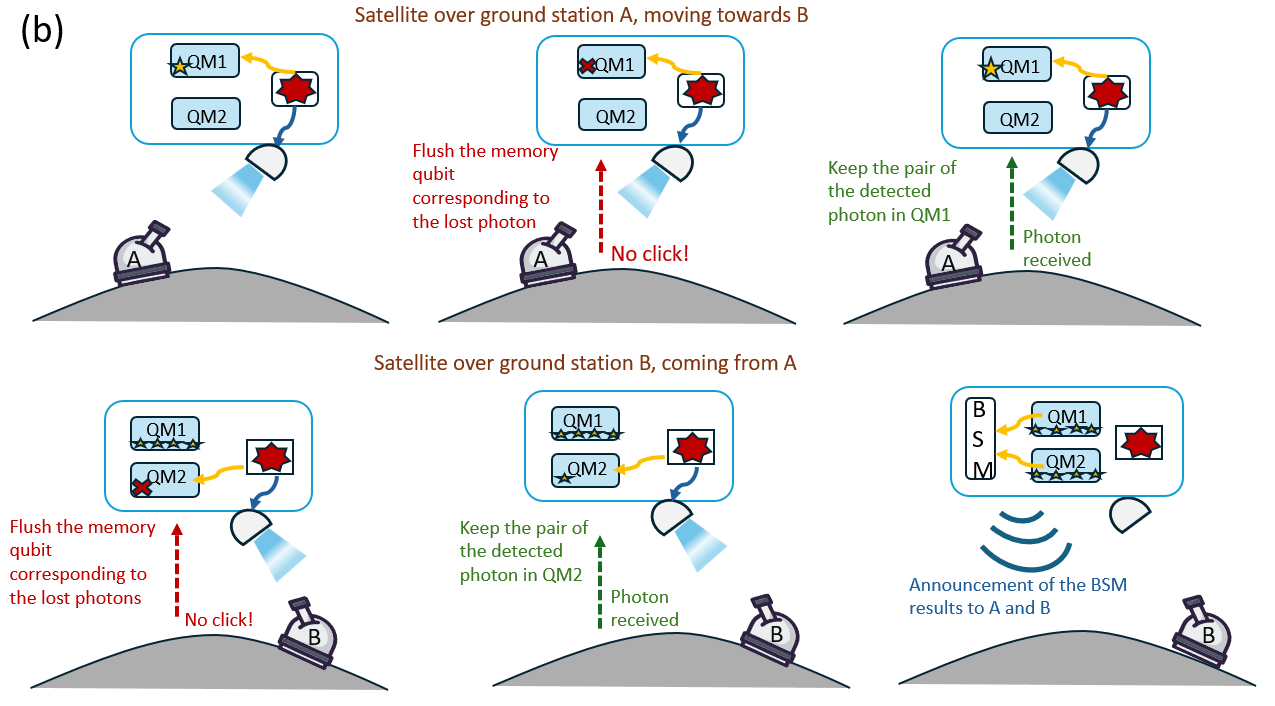}     \end{subfigure}  

    \caption{Entanglement is distributed using very long storage-time memory in a single satellite. (a) One Photon of an entangled pair is stored in the memory (red Square) which is later retrieved and transmitted down-link once the satellite has physically moved to a far-away destination~\cite{simon2017towards, wittig2017concept}. (b) Rates of the single-satellite protocol can be dramatically improved by using a time-delayed repeater protocol using two memories, that result in complete usage of the memory multiplexing capacity~\cite{gundogan2024time}. 
}
    \label{Fig_single_sat}
\end{figure*}


\subsection{Memories in single satellites}

Multiple other architectures are possible using memories and satellites. One such proposal, as described in Fig. \ref{Fig_single_sat}(a), consists of a putting a very long-term memory (storage time in minutes) in satellite which also contains an entangled source \cite{wittig2017concept}. One photon of the entangled pair is transmitted downlink while the other is stored in the memory. If the photon is not received at the ground station, one would flush the memory qubit and try to store a photon in the same qubit again. When the photon is received successfully at the ground station, the qubit will be kept in memory. Now, the next photon can be stored in the next memory qubit and that is how all the memory capacity can be used. 

The satellite then physically moves from one place to another physically and when it reaches the destination the stored photon is retrieved and sent downlink. This protocol would require a very long storage time memory with storage time at least in minutes, preferably in order of an hour as the satellite needs to physically move over a large distance for this to be useful. In this context, rare-earth ion based solid state system has been shown to have coherence times as large as 6 hours \cite{zhong_optically_2015}. More recently, the same system has been engineered to show several improvements including coherence times exceeding 10-hours\cite{wang2025nuclear}. A functioning memory demonstration with more than 1 hour storage time using atomic-frequency combs has also been performed \cite{ma2021one}, although with storage of classical bright pulses.

The long storage time memory also must have high multimode capacity through spatial, temporal or frequency multiplexing for effective transmission rate for several reasons. One is simply that there is a large time delay due to the physical movement of the low-earth satellite to long distances ($\sim$ 20 mins for around 10,000 km). Considering this time delay in transmission, successfully transferring only a few qubits worth of quantum information in a single journey would result in very low transmission rates. Hence, large multiplexing capability is necessary, which can be attained in principle with atomic frequency comb memories with long storage time \cite{seri2019quantum, ma2021one, bonarota2011highly, ma2021one}.

Another important reason for large multiplexing is the loss at each ground link, including the effect of memory efficiency. Consider for example, links with 20 dB loss including memory efficiency, and a memory with 10,000 multiplexing capacity. The first link would be able to fill up the satellite memory, as described previously, with a modest entangled pair generation rate. However, on the second link, only 100 e-bits will be successfully transferred to the ground station. This will mean very low transmission rate and a huge loss of memory capacity.


To solve the above problem, a time-delayed repeater protocol with double quantum memory has recently been invented~\cite{gundogan2024time}, as shown in Fig.~\ref{Fig_single_sat}(b). One photon of an entangled pair is stored, and another is sent down at the ground station just like before, filling up the memory on the satellite.
Then the satellite can fly to its destination station and do the exact same thing using another memory (QM2 in Fig.~\ref{Fig_single_sat}(b)). Then, when all the memory qubits in the other memory are entangled with the second ground station, we can perform entanglement swapping by Bell-state measurements (or C-NOT gates and direct measurements if that can be arranged in the satellite). Such a time-delayed repeater with single satellite and two memories would result in proper use of the maximum quantum memory capacity and hence increase entanglement distribution or QKD rates by several orders of magnitude~\cite{gundogan2024time}. Furthermore, it significantly increases the maximum tolerable loss over its single-memory counterpart~\cite{wittig2017concept} when the finite key effects are taken into account. 

The protocol uses an entangled photon pair source with a rate of 5 MHz, and considers a transmission period of 240 seconds. The analysis includes a memory noise probability of $10^{-3}$, a background count probability of 6.4 $\times$ 10$^{-7}$, and a detector dark count probability of 10$^{-7}$, within a 200 ns temporal window. Memory efficiency is set to 0.6, and detector efficiency to 0.8. \textcolor{black}{These simulation parameters are chosen for illustrative purposes and can be adjusted to explore different scenarios.} The calculations determine the finite key length and secure key rate as a function of average single channel loss and memory dephasing. Results demonstrate that the double memory protocol achieves orders of magnitude higher secure key rates and tolerates greater channel losses than a single memory scheme. For example, with ideal quantum memories, the double memory scheme tolerates channel losses up to 42 dB, a significant improvement over the single memory scheme's limit of 28 dB.

\textcolor{black}{These developments also sparked a complementary line of research: beyond mitigating transmission losses, could the space environment itself enhance the performance of quantum memories? In particular, recent proposals suggest that Bose-Einstein condensate (BEC)-based quantum memories could benefit substantially from microgravity conditions, achieving coherence times and operational stability that surpass their terrestrial counterparts by orders of magnitude~\cite{daros2023proposal}. Such findings open new avenues for leveraging space not only as a transmission channel, but also as an active medium for advancing quantum technologies.}

\section{Satellite-Relay: Just reflect}

A conceptually and technologically simpler approach to achieve global-scale quantum communications has recently been put forward~\cite{goswami2023satellite}.
This protocol uses a satellite chain to simply reflect entangled photons and, crucially, does not require quantum memories or repeater protocols; see Fig. \ref{QN_Review_ASQN_Fig_1}(a).
A chain of satellites in LEO acts as a long-distance optical relay and thereby permits communication without trusted nodes and  entanglement distribution. 
Each satellite uses telescope mirrors to re-focus the light like a lens to compensate light divergence so that the beam can be accommodated by the next satellite. 
The satellites inside the chain move together in the same orbit, greatly simplifying satellite tracking and suppressing channel loss due to pointing errors. 
The chain of satellites in Earth's orbit is described as effectively behaving like a set of lenses on an optical bench, confining photonic qubits faithfully over large distances \cite{goswami2023satellite}.

\subsection{Low loss over global distances}

The proposed architecture, termed All-Satellite Quantum Network (ASQN), minimizes photon loss over long distance in four principal ways.


\begin{figure}[htbp]
    \centering
    \includegraphics[width= \linewidth]{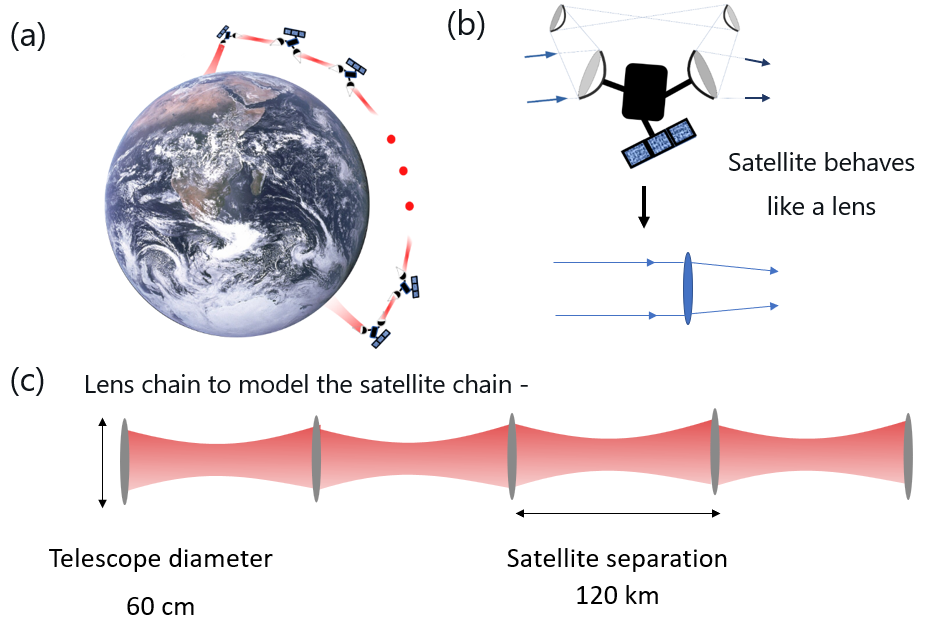}
    \caption{The satellite relay architecture of ASQN is shown, for global scale quantum communication without quantum memory or quantum repeaters. (a) Photons are sent from one side of the globe to another by being reflected from one satellite to another and being guided along the surface of the earth (satellite separation exaggerated). (b) Each satellite has telescopes with curved mirrors that together focuses the light, while also bending it in the direction of the next satellite. Effectively the whole telescope mirror assembly in each satellite behaves like one lens (as shown), without considering the light bending that can't be modeled by a lens. (c) The chain of satellites behave like a set of lenses and contain light beam divergence or diffraction loss indefinitely over very long distance. Diffraction loss at 20,000 km is nearly eliminated (0.67 dB only), while total loss is contained below 30 dB - for satellite telescope diameter of 60 cm, satellite separation of 120 km and 2$\%$ reflection loss at each satellite ~\cite{goswami2023satellite}.
}
    \label{QN_Review_ASQN_Fig_1}
\end{figure}

{\it  1. Elimination of Diffraction Loss using curved satellite mirrors:} Diffraction loss in ASQN is effectively eliminated in the satellite chain through continuous focusing by curved telescope mirrors. 
That is, the telescope assembly at each satellite creates an effective "satellite lens” to re-focus light. The telescope assembly can be either a four-mirror (conjugate beam) system  or a two mirror system (as in Fig. \ref{QN_Review_ASQN_Fig_1}(b) and (c)). Both systems can focus input light like a lens (see Fig. \ref{QN_Review_ASQN_Fig_1}(d)).
The chain of satellite lenses mitigate diffraction over long distance as illustrated in Fig. \ref{QN_Review_ASQN_Fig_1}(e). The diffraction loss is simulated numerically by propagating the beam from one lens to the next, providing a thin lens phase shift at the next lens, and then using an aperture to truncate the beam according to the lens size.
Simulations confirm the near-complete elimination of diffraction loss (only 0.67 dB) even at global distances of 20,000 km - for lens (satellite) separation of 120 km, lens (telescope) diameter of 60 cm and 800 nm wavelength. Beam size at the satellite is chosen as 17.48 cm, so that beam's Rayleigh range exactly matches satellite separation. ``Satellite Lens" focal length of 60 km focuses the divergent beam exactly at the midpoint (at 60 km) between two satellites, due to focal shift~\cite{li1981focal, renk_basics_2012}.
The diffraction loss has been found to be robust against the effect of lens (satellite) position errors and focal length errors as well, as discussed later in this section. This indirectly verified effects of different satellite motions which isn't simulated directly as a stationary set of lenses is considered. Apart from focusing, the telescope mirrors slightly bend the light beam to the next satellite, eventually directing it along the Earth's curvature.

\begin{figure}[t]
    \centering
     \centering        \includegraphics[width= 0.45\textwidth]{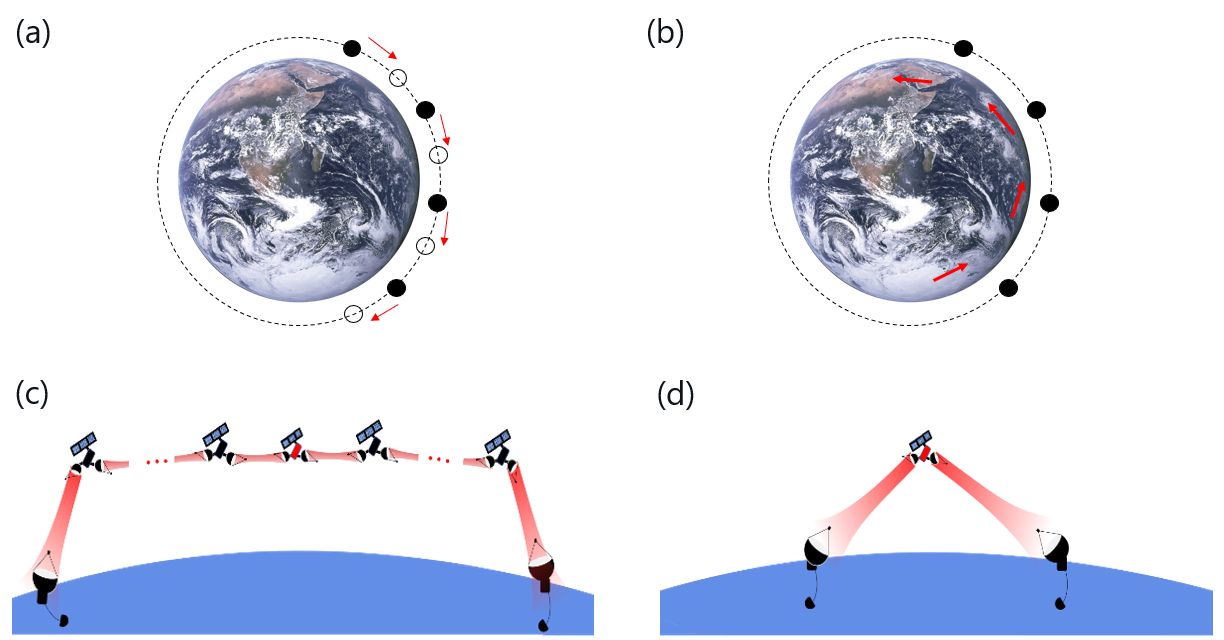}

    \caption{Satellite tracking loss in ASQN~\cite{goswami2023satellite} is nearly eliminated within the satellite chain as shown in (a)-(b). (a) The Satellite chain moves synchronously (in the direction of red arrows). The filled and empty circles shows the initial and later position of the satellites. (b) In the reference frame of one satellite - all other satellites are stationary and the Earth moves below. So, almost no dynamic tracking (i.e. pointing or point-ahead) is needed within the satellite chain, except due to errors in  satellite positioning. The satellites only need to be aligned for tranmission along the chain. (c)-(d) Ground link loss is much less in satellite-relay compared to single satellite transmission. (c) The chain of satellites enables near-vertical tranmission to ground, while a more oblique incidence is necessary in for single satellites as shown in (d).
}
    \label{QN_Review_ASQN_Fig_2}
\end{figure}

{\it 2.	Synchronous Orbits Minimize Tracking:} Tracking error is also nearly absent within the satellite chain in ASQN because satellites inside the chain are co-moving in the same orbit (Fig. \ref{QN_Review_ASQN_Fig_2}(a)). 
Hence, the satellites are stationary relative to each other in their frame of reference, as depicted in Fig. \ref{QN_Review_ASQN_Fig_2}(b). 
Thus, mostly just light alignment is necessary in the chain along with very small residual tracking for orbital deviations and other errors.
Tracking is only required at the up- and down-links or between independent satellite chains in different orbits, as discussed in the next section.

{\it 3.	Only Reflection at each Satellite:} Loss at each satellite (e.g. component absorption loss) is the only exponentially scaling loss in ASQN. 
It is minimized by only allowing mirrors in light path, at least in most satellites, because of the inherently low loss in mirrors. If the two large mirrors are metal mirrors with as much as 1$\%$ reflectivity loss, there would be a resonable 15 dB total reflection loss at 20,000 km. If smaller front mirrors are present (like in Fig. \ref{QN_Review_ASQN_Fig_1}(b)), they can be ultra-high reflectivity Bragg mirrors with negligible loss ($<$ 0.0001$\%$). If Bragg mirrors can be used for the two larger mirrors as well~\cite{pinard2017mirrors}, even the mirror loss can be completely eliminated. 

{\it 4.	Vertical Ground Link:} The preceding discussion focused on minimizing losses within the satellite chain, which constitutes the longest-distances in the communication channel. 
Indeed, ASQN also involves ground-to-satellite and satellite-to-ground links like all other satellite protocols, which experience diffraction, attenuation, and turbulence losses. 
However, ASQN utilizes near-vertical ground transmission links, reducing losses associated with grazing or oblique incidence commonly encountered in single LEO satellite as shown in Fig. \ref{QN_Review_ASQN_Fig_3}(c). 
Hence, diffraction and atmospheric losses associated with ground link transmission in ASQN (even for a 20,000 km transmission) will be significantly less than that of a single satellite like Micius, as depicted in Fig. \ref{QN_Review_ASQN_Fig_3}(d) , if a similar-sized telescope and the same elevation is considered. Total diffraction loss including ground link is only 5 dB in ASQN as 60 cm telescopes and vertical links are used. Satellites are assumed to be at 500 km orbit with 1.2 m receiving telescopes in ground.
Hence, in ASQN photon loss is low both in short and long distances, compared to other protocols involving oblique satellite transmission.

Considering all the above factors and further losses like that from atomspheric transmission, as well as the effect of errors,  the total loss in ASQN for entanglement distribution is contained below 30 dB for a 20,000 km optical path length in the satellite chain.
Hence, a 1 GHz source would enable a 1 MHz entanglement distribution rate even at 20,000 km.
This calculation assumes a satellite based source, 60 cm-diameter satellite telescopes, 120 km satellite separations, and 2$\%$ reflection loss at each satellite, as stated above. Effect of errors on diffraction showed a 12 km (10$\%$) error in satellite separation, 0.6 cm (2$\%$) error in telescope lateral position and 3 km (5$\%$) in focal length error contribute to only around 5.7 dB extra diffraction loss at 20,000 km. This directly verify robustness against focusing errors and indirectly verify effects of satellite motion and orbital deviations inside the satellite chain that leads to residual tracking requirements, as explained in~\cite{goswami2023satellite}. 
The different factors of loss, in entanglement distribution from a satellite-based source, are summarized in the following table ( for parameters mentioned earlier in the text).
~\\

\begin{tabular}{ |p{6.5cm}||p{1.5cm}| }
 \hline
 \multicolumn{2}{|c|}{Loss due to different factors} \\
 \hline
 Diffraction loss within the satellite chain & 0.67 dB \\ 
 Total diffraction loss including ground link & 5 dB\\
 Reflection loss (2$\%$ at each satellite) & 15 dB\\
 Other losses - atmospheric attenuation and loss due to errors and inefficiencies    & 10 dB\\
 \hline
 Total entanglement distribution loss at 20,000 km &   30 dB \\
 \hline
\end{tabular}
~\\



\subsection{Background and Protocols}

As mentioned, satellite chains for trusted-node QKD has been proposed \cite{pirandola2021satellite, noauthor_china_2024}.
Also, some optical relay designs have been proposed before for quantum communication using drones and satellites~\cite{liu2021optical, aspelmeyer2003long}. 
However, only high-orbit reflectors were considered for satellites, which led to a high degree of diffraction for photons because of the large distances covered through high orbits and no focusing of light. 
Nevertheless, this approach requires only a few reflectors and allows a wide area of collection.
We note that several experiments have been performed using retro-reflectors in high orbits like geo-stationary orbits to experimentally ascertain channel loss ~\cite{bedington_progress_2017, vallone_experimental_2015,calderaro2018towards}. 
ASQN differs from previous high-orbit studies in that its low-loss telescope mirror assemblies mitigate enormous diffraction loss through continuous focusing and limits other losses, allowing the distribution of quantum-level light signals over very long distances.

There has been consistent effort in creating inter-satellite laser links for classical communication~\cite{guelman2004acquisition, lakshmi2008inter, chaudhry2022temporary}.  
In these implementations however, all the satellites are moving at high speeds with respect to each other and the satellites are generally not equipped with large telescopes, since light is just detected at the next satellite and classical ground link  rely on radio frequency communication. 
Recently, nearly continuous inter-satellite laser links with more 100 GB/s capacity has been achieved with more than 5000 satellites by the Starlink internet service~\cite{brashears2024achieving,chaudhry2021laser}.

ASQN can accommodate different quantum communication protocol. Till now, we  discussed the entanglement distribution protocol with the entanglement source on a satellite, sending photons down to the Earth via two downlinks. 
Another protocol, termed as “Qubit Transmission” was proposed in ASQN. It is a prepare and mesaure scheme where weak laser pulses are sent in one way transmission from a ground source to a ground detector through the satellite-relay. This scheme can perform QKD using the BB84 protocol~\cite{bennet1984quantum}.
Placing both the source and detector on the ground offers significant advantages in terms of space, energy, access to advanced technologies like cryogenics, and qubit multiplexing capabilities. Most importantly, both the source and detector can be changed to create different protocols and experimentation.

Qubit transmission will have one uplink and one downlink transmission, facing uplink atmospheric turbulence. 
Uplink turbulence distorts, fragments and enlarge the beam and only a small portion of the beam is caught in the first satellite.
However, after the uplink to satellite - ASQN's satellite lenses effectively control further spread of beam divergence along the chain from the distorted beam (see Fig. 6 of~\cite{goswami2023satellite}). 
Simulations using multiple phase screens~\cite{kolmogorov1941local, goswami2023satellite} confirmed that  diffraction loss is limited after the first uplink. Though the distorted beam contain different order modes of light, the lens systems confine fundamental/low-order modes, ensuring only 2 dB ($\sim$ 35$\%$) extra diffraction loss at 20,000 km (negligible compared to the initial 22 dB uplink loss at 500 km orbit). This can be understood as the distorted beam containing 35$\%$ higher order modes that can't be confined.
The impact of transmitting the distorted beam is seen indirectly though, in requiring closer satellite separation of 80 km in the simulation - instead of the 120 km separation needed for entanglement distribution (without uplink).

In ASQN, worldwide coverage can be enabled using nearly orthogonal satellite chains covering the entire globe. The bulk of the transmission will be along the satellite chains where satellite tracking will be minimal due to synchronous orbits. However, to connect two arbitrary points, one connection between two separate chains will be necessary where dynamic satellite tracking will be needed. Such tracking has already been achieved in classical satellite-to-satellite links \cite{brashears2024achieving,chaudhry2021laser}. If a worldwide network is constructed, any point on the ground should always have one (ideally multiple) satellite in its view. Hence, the network in ASQN will not be limited in satellite flyby time. As one satellite passes further, one can always point at the next satellite and establish nearly continuous quantum communication. 
Beyond flyby, the use of multiple chains will allow optimizing the network capacity, which is an open research question in quantum communication.
This question has been very recently explored for ASQN connected with ground-based fiber networks~\cite{gu2025quesat}.

\subsection{Challenges}

Along with introducing the protocol, a feasibility study of ASQN has been conducted in~\cite{goswami2023satellite}, exploring various aspects such as the use of vortex beams for on-axis telescope setups, alignment and tracking procedures, the effects of space on mirror performance, different forms of wavefront aberration, different qubit types, and focal length adjustments for the satellite lenses. There are of course more effects to analyze like temperature dependence, source etendue effect~\cite{welford1978optics, markvart2007thermodynamics}, detailed simulation of satellite motion and acquiring, pointing and tracking (APT) process, possible small effects in optical near-field and more such detailed engineering concerns. 
The recent rapid progress in space industry~\cite{palmer2021spacex, heldmann2022mission, brashears2024achieving, chaudhry2021laser, henri2020oneweb,osoro2021techno} provides the backbone to launch (possibly even affordably soon) a large network of satellites required for ASQN, in which one chain spanning 20,000 km will itself require about 160 ($\sim$ 20,000/120) satellites. Overall, a comprehensive diffraction analysis demonstrated the effectiveness and robustness of ASQN showing the practicality of a satellite-relay while acknowledging the engineering challenges associated with building and deploying such a network, particularly wavefront aberration control in mirror surface design and residual satellite tracking requirements over the long chain.



\subsection{Ground-based Relay: Vacuum beam-guides}

A similar, but ground-based, relay proposal has recently emerged~\cite{huang2024vacuum}, using a series of lenses within vacuum tubes. These are actual lenses, in contrast to the mirror setup working as an effective lens in ASQN.
The lenses periodically focus light to transmit quantum information over a long distance. 
This approach, termed vacuum beam guides, directly employs lenses spaced 4 km apart inside vacuum tubes, and is partially inspired by Laser Interferometer Gravitational-Wave Observatory (LIGO) infrastructure~\cite{abbott2009ligo}. 
Numerical modeling considered loss due to scattering from lenses, absorption loss from residual gas in the vacuum tube, and lens alignment errors - promising only $10^{-4}$ dB/km loss.
This proposal avoids dealing with atmospheric up- and down-links, in contrast to satellites in ASQN. 

However, vacuum beam guides have their own challenges. 
For instance, lenses cannot bend light passing through their center, and additional optical elements like mirrors or prisms (with additional losses) would be necessary to accommodate the Earth's curvature. 
However, mirror assemblies like that used in ASQN~\cite{goswami2023satellite} can mitigate this.
One other issue is lens vibration from ground, which was taken into account in simulations, but can vary depending on the local environment.
Focal length fluctuation of the large (4 km) focal length lenses due to thermal variations is another concern. 
A broader and more important issue concerns the practicalities of constructing and maintaining the numerous vacuum beam guides for world-wide communications, including taking into account the varying geography and terrains as well as political and environmental considerations.
In contrast, satellites are much more established, including their manufacturing and operation as well as legal and political considerations. 
However, given there is no practical solution to achieve long-distance quantum communication currently – all avenues and ideas should be pursued enthusiastically until a clear winner is established.

\section{Combination of Satellite-relay and quantum memories}


\begin{figure}[htbp]
    \centering
     \begin{subfigure}
     \centering        \includegraphics[width=0.5\textwidth]{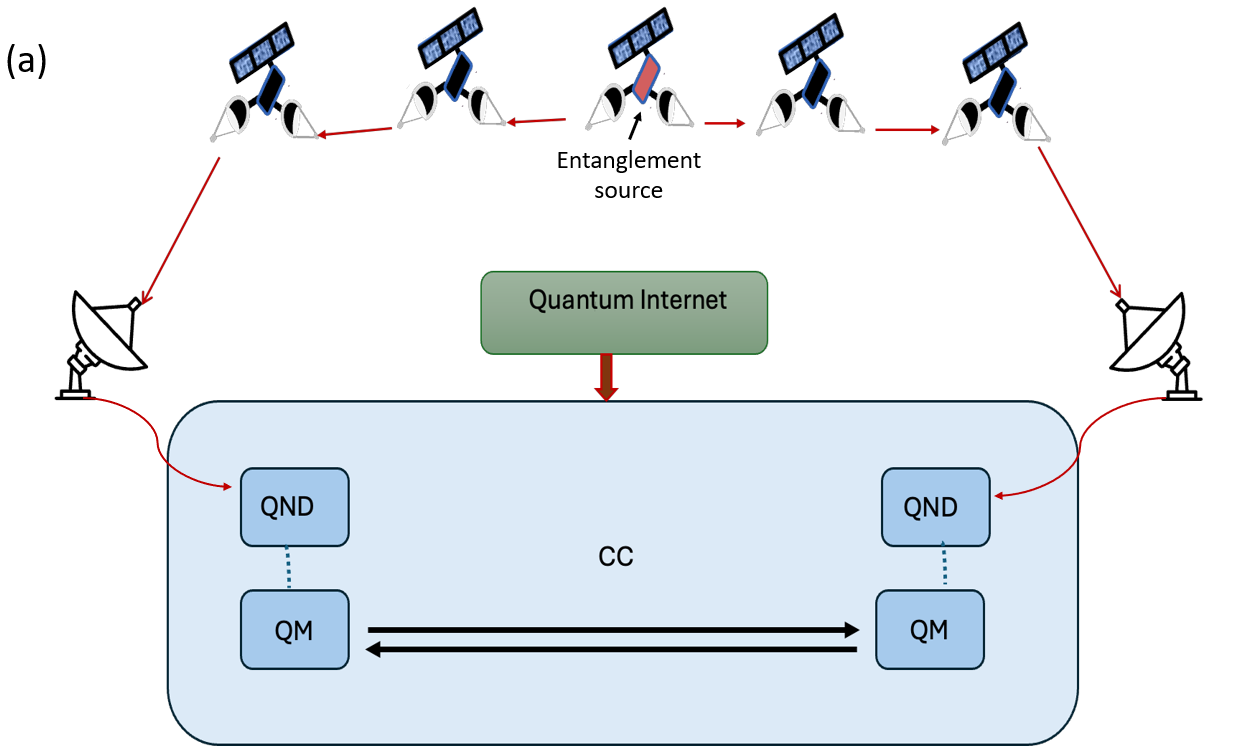}     \end{subfigure}
     \begin{subfigure}
     \centering        \includegraphics[width=0.5\textwidth]{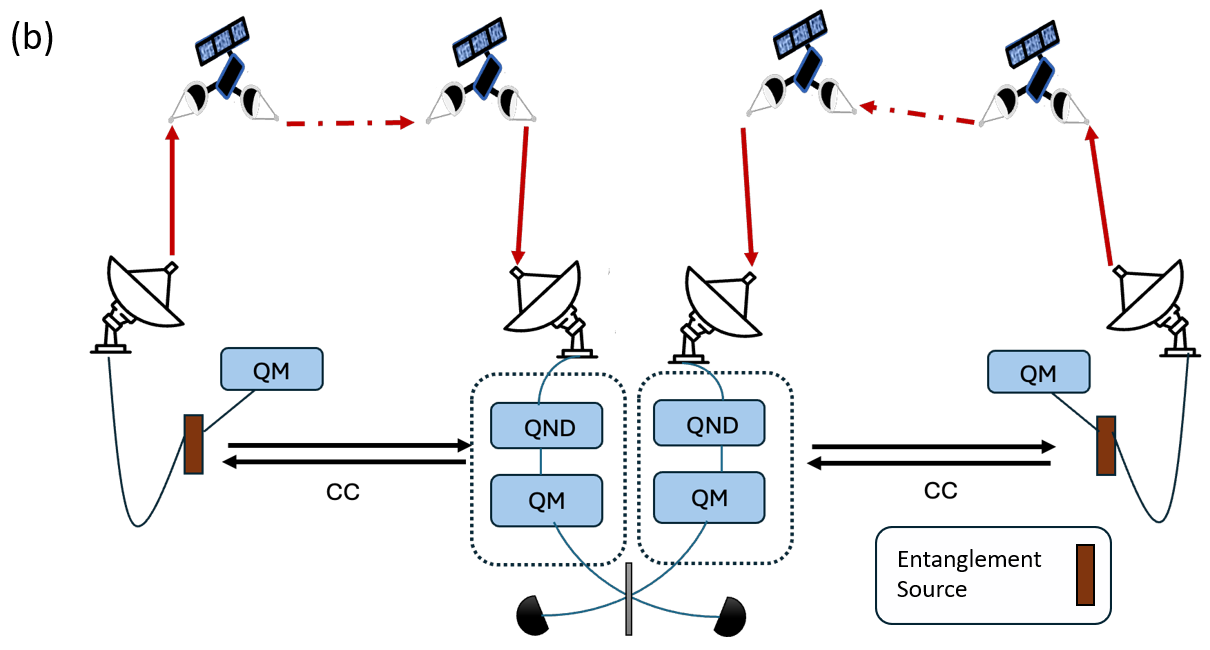}     \end{subfigure}   

    \caption{Combination of satellite-relay and quantum memories. (a) Only two Quantum memories (QM) can be incorporated only at the end to create quantum internet capabilities in satellite-relay architecture of ASQN~\cite{goswami2023satellite}. However, that will require Quantum non-demolition (QND) detectors. (b) Repeater over relay - with two detectors at the middle, QND detectors are not essential anymore. Quantum internet capabilities are achieved with higher rates. Note that the QM and QND in the middle are optional to increase rates.  
}
    \label{QN_Review_ASQN_Fig_3}
\end{figure}


Quantum memories will likely be required in quantum networks to achieve advanced functionalities \cite{wehner_quantum_2018}.
Some of the stringent requirements on memory performance, such as efficiency and multiplexing capacity, that are required for conventional ground-based repeaters or other memory-satellite proposals can be relaxed though due to satellite-relay's elimination of diffraction loss~\cite{goswami2023satellite}. 
Especially, memory efficiency will not be a big factor as only a few memories (e.g., two) are needed. Multiplexing capacity needed will also be more relaxed.
The primary reasons for incorporating quantum memories for quantum network are due to heralding and causality as well as further loss avoidance.
Quantum memories store entangled photons while waiting for a classical signal that indicates which specific entangled qubit has been heralded to another location in multiplexed communications (heralding) or what the outcome of a measurement was on an entangled qubit at a distant location (causality).
Several operations, like quantum teleportation,
require quantum operations that are conditioned on results of previously performed measurements due to causiality.
Loss reduction due to repeaters is especially relevant for up-link transmission. For downlink entanglement distribution rates can be increased much further too by using satellite repeaters nodes in combination with the relay. In the space-based satellite relay, loss in inter-satellite links of 2,000 km is more than 15 dB~\cite{gundogan_proposal_2021}. Using multiple relay nodes inside each inter-satellite link will decrease the loss drastically. For example, 2000 km space transmission loss in ASQN is less than 2 dB with about 0.1 dB diffraction loss and 1.5 dB reflection loss. A relay-repeater combination with such low loss links can achieve higher entanglement distribution rates than what both relay and repeaters can achieve individually. However, to achieve very high rates large multiplexing capabilities (time or frequency) will be needed for quantum memories along with long storage time~\cite{goswami2023satellite}.




Different strategies for incorporating quantum memories into ASQN has been proposed (see Fig. \ref{QN_Review_ASQN_Fig_3})~\cite{goswami2023satellite}. 
For instance, QND can identify and quantum memories can store photons distributed by ASQN, e.g. for interfacing with a ground-based networks, quantum computers, or sensors (see Fig. \ref{QN_Review_ASQN_Fig_3}(a)). 
Alternatively, a repeater protocol (with or without QND) can be performed using ground-based entangled sources but replacing the terrestrial channels with space transmission (Fig. \ref{QN_Review_ASQN_Fig_3}(b)). 
Clearly, there are open research questions how satellite-relays can be used in conjunction with memories, repeater protocols and other forthcoming quantum technologies.

Importantly, the above schemes emphasize the flexibility of satellite-relay in accommodating both relay and repeater functionalities. While relay's primary advantage lies in eliminating the need for quantum memories in the initial stages of building a global quantum network, it can be seamlessly integrated with quantum repeaters when necessary. This hybrid approach allows for a staged development of the quantum internet, starting with simpler functionalities and gradually incorporating more complex ones as quantum memory technology matures.

\section{Conclusion}

Global quantum communication networks face the formidable challenge of transmitting quantum information over long, lossy, channels. 
While ground-based quantum repeaters offer a potential solution, their reliance on high-performance quantum memories remains a significant hurdle. 
Satellite-based approaches, exemplified by the Micius satellite achieving entanglement distribution over 1200 km, have demonstrated remarkable success in overcoming loss limitations.
In the tailwinds of this demonstration, we reviewed several long-distance quantum communications proposals using satellites: either with satellite-based repeaters, satellites equipped with long-lived quantum memories, or satellite-relay protocols where photons are reflected from one satellite to another. 
Such approaches benefit from the rapid progress in the space industry, including the advent of rapidly reusable mega-rockets~\cite{elvis2023accelerating, palmer2021spacex, heldmann2022mission} and deployment of large-scale classical-internet satellite constellations reducing launch costs~\cite{brashears2024achieving,chaudhry2021laser, henri2020oneweb, osoro2021techno}.
The eventual architecture for a quantum internet may even be a combination of different approaches, whether its goal is  to achieve global scale QKD, diverse networked quantum technologies building towards a quantum internet, to realize fundamental science, or beyond.

\begin{acknowledgments}
Authors acknowledges Aephraim M. Steinberg for fruitful discussions.
M.M. and J.R.L. acknowledge support from Boeing.
J.S.S. and D.K.L.O would like to acknowledge the support of the EPSRC Quantum Technology Hub in Quantum Communications (EP/T001011/1). This work was supported by the EPSRC International Network in Space Quantum Technologies INSQT (EP/W027011/1). 
D.K.L.O. is supported by the EPSRC Integrated Quantum Networks Hub (EP/Z533208/1).  
M.K. and M.G. acknowledge the support from the DLR through funds provided by the Federal Ministry for Economic Affairs and Climate Action (Bundesministerium für Wirtschaft und Klimaschutz, BMWK) under Grants No. 50WM1958, 50WM2055 and 50WM2347.  
M.G. further acknowledges funding from the European Union’s Horizon 2020 research and innovation programme under the Marie Sklodowska-Curie grant agreement No. 894590 and Einstein Foundation Berlin for support. 
Y.C.C. and H.H.J. acknowledge support from the National Science and Technology Council (NSTC), Taiwan, under the Grants Nos. 112-2112-M-001-079-MY3 and NSTC-112-2119-M-001-007, from Academia Sinica under Grant AS-CDA-113-M04 and are also grateful for support from TG 1.2 of NCTS at NTU. 
C.S. acknowledges the Natural Sciences and Engineering Research Council of Canada for its Alliance Quantum Consortia grants ARAQNE and QUINT, and the National Research Council for its High-Throughput Secure Networks challenge program.
\end{acknowledgments}




%

\end{document}